\documentclass[twocolumn]{revtex4-1}

\usepackage{lipsum}
\usepackage{bm}
\usepackage{color}
\usepackage{graphicx}
\usepackage{amsmath}
\usepackage{amssymb}

\begin{document}

\title{\Large{Discovery of entanglement generation by elastic collision\\ to realise the original Einstein-Podolsky-Rosen thought experiment}} 

\author{Roman Schnabel} 
\affiliation{
$^{1}$ Institut f\"ur Quantenphysik \& Zentrum f\"ur Optische Quantentechnologien, Universit\"at Hamburg, Luruper Chaussee 149, 22761 Hamburg, Germany}

\email{roman.schnabel@uni-hamburg.de}
\date{May 14th, 2025, corrected illustrations on June 19th, 2025}
\maketitle

\section*{Abstract} \vspace{-3mm}
{\bf 
The amazing quantum effect of `entanglement' was discovered in the 1935 thought experiment by Albert Einstein, Boris Podolsky and Nathan Rosen (`EPR') \cite{Einstein1935}. The ensuing research opened up fundamental questions and led to experiments that proved that quantum theory cannot be completed by local hidden variables \cite{Freedman1972,Aspect1981,Pan2000}. Remarkably, EPR did not discuss how to create the entanglement in their thought experiment. Here I add this part. What is required in the original EPR thought experiment is a simple elastic particle collision, an unbalanced mass ratio of e.g.~1:3 and initial states that are position and momentum squeezed, respectively. In the limiting case of infinite squeeze factors, the measurement of the position or momentum of one particle allows an absolutely precise conclusion to be drawn about the value of the same quantity of the other particle. The EPR idea has never been tested in this way. I outline a way to do this.
}

\vspace{-2mm}
\section*{Introduction} 
\vspace{-2mm}
In their seminal 1935 paper \cite{Einstein1935}, Einstein, Podolsky and Rosen presented a thought experiment with two systems that had interacted with each other in the past but no longer do. The past interaction had `entangled' the systems, a term introduced by Erwin Schr\"odinger in his response \cite{Schroedinger1935}. EPR were finally specific in the presentation of their thought experiment. They considered the entanglement of the momenta and positions (coordinates) of two particles, which I will refer to here as `A' and `B'. For a statistical analysis, the entanglement was prepared many times in exactly the same way. The successive measurements on particle A (its location $\hat x_{A,i}$ or its momentum $\hat p_{A,j}$) as well as those on particle B ($\hat x_{B,i}$ or $\hat p_{B,j}$) nevertheless resulted in different values due to quantum uncertainty.

The core of the EPR thought experiment was the discovery of correlations \emph{within} the spread of the uncertainty distributions. Every two simultaneous position measurements ($x_{A,i}$ and $x_{B,i}$) did provide varying values but they were always mutually identical. Two simultaneous momentum measurements ($p_{A,j}$ and $p_{B,j}$) also provided varying values, but always had a sum of precisely zero. 
The fact that any quantum uncertainty disappears in the \emph{relative} measurements on a pair has led EPR to question whether quantum theory is complete \cite{Einstein1935}. Many other quantum physicists did not question this. Schr\"odinger saw nevertheless a paradox in the EPR thought experiment \cite{Schroedinger1935}.

EPR did not describe in their thought experiment how the position/momentum entanglement of two particles could be realised.
Starting in the 1970s, there are now a large number of EPR experiments that realise the EPR paradox with different observables and different quantum systems. These include entangled systems of definite photon numbers \cite{Freedman1972,Aspect1982,Pan2000}, followed by conceptionally similar experiments with the occupation numbers of internal states of trapped ions \cite{Sackett2000}, of two atoms and a cavity mode \cite{Rauschenbeutel2000}, of the stretch modes of two separated atomic mechanical oscillators \cite{Jost2009}, of electron spin oscillators in defects of two separated crystals \cite{Hensen2015}, and of phonon number excitations of two artificially engineered mechanical oscillators \cite{Riedinger2018}.
A second kind of EPR experiments has used \emph{indefinite} numbers of quanta and produced entanglement of the position-like and momentum-like observables, in particular the amplitude and phase quadratures amplitudes $\hat X$ and $\hat Y$. Their continuous-variable probability density distributions of eigen values obey the Heisenberg uncertainty relation $\Delta\!^2\!\hat X \cdot \Delta\!^2 \hat Y \geq 1/16$, where the $\Delta\!^2$ denote variances. Also here, the first demonstrated systems were optical, namely laser beams having well-defined optical frequencies, polarisations, and transverse modes \cite{Ou1992,Bowen2003}. 
Another example is EPR experiments with the transverse position and momentum of optical fields in the context of imaging \cite{Aspden2013,Kumar2021}. As an example of systems with mass, the position- and momentum-like projections of the collective spin of atomic clouds were entangled, e.g.~clouds of about $10^{12}$ caesium atoms \cite{Julsgaard2001} or about $10^{4}$ rubidium atoms \cite{Peise2015}.

Here I present the previously unknown interaction of how two free particles get entangled with respect to their real positions and momenta, namely those of their centre of mass motion, to realise the source for the original EPR thought experiment. It is an elastic collision in one dimension, where the bodies must have unequal mass and are initially in position- and momentum-uncertainty-squeezed Gaussian states, respectively. I show that the entangling process is a natural consequence of the existence of quantum uncertainty and conservation of energy and momentum. I use a semi-classical approach that neglects the interference of the uncertainty ranges (during the collision). I consider this to be well-founded, as significant interference only occurs when very similar wave functions overlap. In the case considered here, neither the masses of the systems nor their quantum states are the same. The {\it measurement} of the positions and momenta only takes place when the wave functions are clearly separated again by the kinetics, so that my model is not subject to a semi-classical approximation during the measurement. Note that the (Bargmann) mass-superselection rule does not apply, because the masses of the particles are no dynamical variables \cite{Giulini1996}.\\
I propose to put the EPR thought experiment into reality with precisely those systems and system observables originally discussed by EPR using an ensemble of a large number of identically prepared position$/$momen\-tum entangled pairs of freely propagating atoms or ions.\\

\begin{figure*}[ht!!!!!!!!!!!!!!!!]
 \center 
     \vspace{0mm}
    \includegraphics[width=16cm]{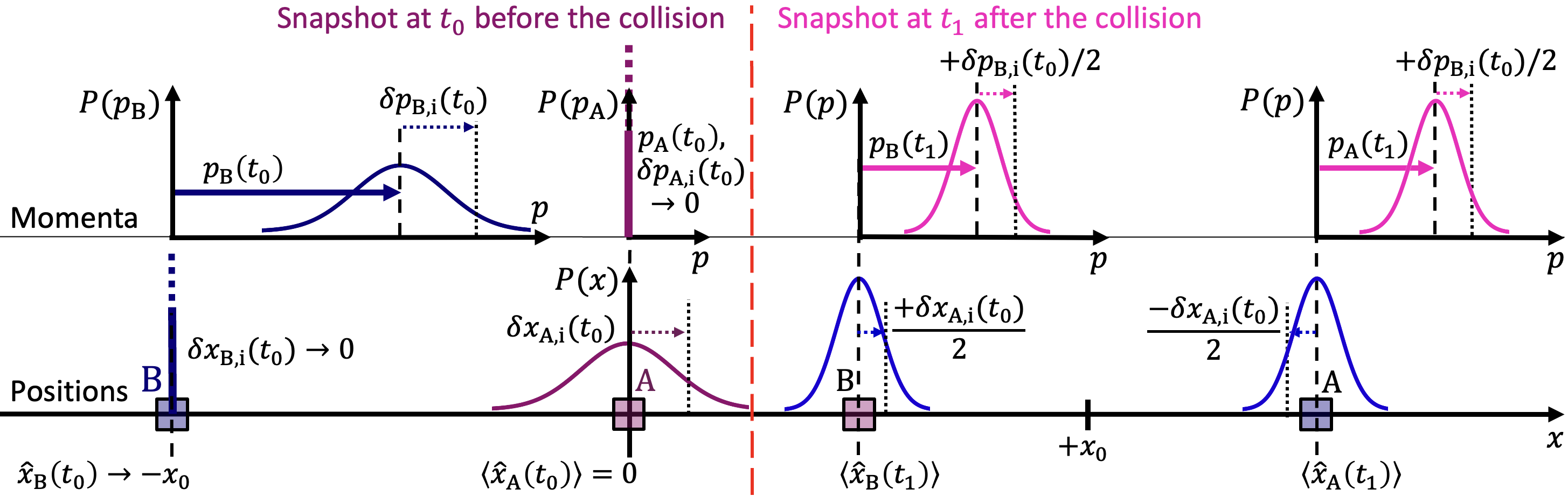}
    \vspace{-1mm}
    \caption{{\bf EPR entanglement from elastic collision} -- Before the collision, at time $t_0$ (left half), particle A with mass $m_{\rm A}$ rests at $\langle x_{\rm A}(t_0) \rangle = 0$ with a large Gaussian position uncertainty and negligible, strongly squeezed momentum uncertainty. The entanglement is produced by a single collision with particle B having the mass $m_{\rm B} = 3 m_{\rm A}$, a high momentum $\langle \hat p_{\rm B}(t_0) \rangle  \gg \Delta \hat p_{\rm B} (t_0) \gg 0$, and a position $\hat x_{\rm B}(t_0) = -x_0$ with negligible, strongly squeezed position uncertainty. After the collision, at time $t_1$ (right half), measurements are performed. The two masses have the same momentum $\langle \hat p_{\rm B}(t_1) \rangle \!=\! \langle \hat p_{\rm A}(t_1) \rangle$ due to the mass ratio 1:3 and the conservations of momentum and energy. The momenta are even identical for any individual ensemble measurement `i' since the momentum uncertainties (almost) exclusively originate from particle B. The measured values $\langle \hat p_{\rm A,B}(t_1) \rangle + \delta p_{\rm A,B,i}(t_1)$ (vertical dashed lines in the Gaussian distributions top right) are always identical, i.e.~the differential values show no quantum uncertainty. The momentum uncertainties of the two particles are quantum correlated.
    The initial position uncertainty $\Delta \hat x_{\rm A}(t_0) \gg 0$ gets also distributed onto both particles. The right half of the Gaussian uncertainty corresponds to a statistically later collision, which results in a later establishment of new velocities. The collision halves the velocity of mass B, and mass A is accelerated to $3/2$ of the initial velocity of B. The position uncertainty of A is therefore mirrored at its centre line and compressed by a factor of $1/2$ due to A's uncertain initial position, see Eq.\,{eq:11}, while B takes over the other half of A's position uncertainty without a change of sign (see supplement). In conclusion, the position uncertainties of particles A and B after the collision are quantum anti-correlated. 
  \emph{By measuring either A or B we are in a position to predict with certainty, and without in any way disturbing the second system either the value of the quantity $[\,x\,]$ or the value of the quantity $[\,p\,]$.} My complemented version of the EPR thought experiment makes obvious that the description by the wave function is complete. Hidden variables are not motivated by the EPR thought experiment.
    }
    \label{fig:1}
\end{figure*}

{\it Gaussian quantum uncertainties --}
The emergence of motional (position$/$momen\-tum) EPR entanglement of two particles in Gaussian quantum states through an elastic collision is illustrated in Fig.\,\ref{fig:1}. The equivalent \emph{phase space} description is presented by the figure of the supplementary information.

States with Gaussian quantum uncertainties can minimize Heisenberg's uncertainty relation \cite{Heisenberg1927,Kennard1927,Weyl1927,Robertson1929}. 
The most prominent pair of non-commuting observables whose eigen value spectrums can show a Gaussian distribution are the position $\hat x$ and the momentum $\hat p$, the two observables of motion. 
With $\Delta^{\!2} \hat x$ and $\Delta^{\!2} \hat p$ being the variances of their quantum uncertainties, Heisenberg's uncertainty relation reads
\begin{equation}
\Delta^{\!2} \hat x \cdot \Delta^{\!2} \hat p \geq \frac{\hbar^2}{4} \, ,
\label{eq:1}
\end{equation}
where $\hbar$ is the reduced Planck constant. 

Two systems `A' and `B', both of which must generally obey inequality (\ref{eq:1}), are in a position$/$moment entangled state, if the measurement values on the individual systems reveal correlations tighter than the minimum uncertainty product in inequality (\ref{eq:1}). A sufficient and necessary criterion for the most 
generic form of Gaussian entanglement, which is also called `inseparability', is given in \cite{Duan2000,Simon2000}.  
If the measurement values from one system allow for the inference of the values of the same observable from the second system to better than the second system's minimum uncertainty product $\hbar^2 / 4$, then we speak about `EPR entanglement' \cite{Reid1989,Bowen2003,Haendchen2012,Schnabel2017}. 
EPR entangled systems are always also inseparable, while the converse of this statement does not apply \cite{Bowen2003,Haendchen2012}. 
EPR entanglement corresponds to the correlations found in the original thought experiment of Einstein, Podolsky and Rosen \cite{Einstein1935}.\\

{\it Squeezed motional states of a particle --}
A particle in a harmonic potential being in its motional ground state has position and momentum expectation values of $\langle \hat x \rangle = \langle \hat p \rangle =0$, defined with respect to the mini\-mum of the trapping potential. Precise measurements performed on identical such particles, however, map out the respective quantum uncertainties around zero. They have a continuous Gaussian spectrum of `eigen values'.
The variances of the measured eigen values read
\begin{equation}
\Delta^{\!2} \hat x = \frac{\hbar}{2m \Omega} \, , \hspace{6mm} \Delta^{\!2} \hat p = \frac{\hbar m \Omega}{2} \, ,
\label{eq:2}
\end{equation}
where $\Omega$ is the trapping angular frequency. 
If the trapping frequency is quickly reduced, its state becomes position squeezed:
The (old) position uncertainty is smaller than the ground state's position uncertainty in the new potential.
Such a state is represented by a vertically aligned ellipse in the position/momentum phase space, see the left ellipse in the figure of the supplementary information.
If the trapping frequency is quickly increased, the state becomes momentum squeezed, see the lower ellipse in the same figure.
If the trapping potential turns to a flat potential, the particle gradually increases its squeezed position uncertainty towards infinity and reduces its anti-squeezed momentum uncertainty to zero. After an infinitely long time, the state of the particle reaches the new ground state with infinitely extended position uncertainty and precise momentum. This is the well-known `free evolution of a Gaussian wave packet' that can be found in textbooks on quantum physics. 
Fig.\,\ref{fig:1} and the figure of the supplementary information, however, show entanglement generation on a time scale that is much shorter than the time scale of free evolution and the latter does not need to be considered.

\vspace{-2mm}
\section*{Results}
\vspace{-7mm}
~\\
{\bf State preparation before the entangling collision}\\
Fig.\,\ref{fig:1} illustrates the positions and momenta of two particles `A' and `B' at fixed state-preparation time $t_0$ and at fixed measuring time $t_1$.
The entangling collision happens at time $t_{\rm coll}$ ($t_0 \!<\! t_{\rm coll} \!<\! t_1$) at the position of resting particle A ($\langle \hat x_{\rm A} (t_0) \rangle = 0$).
The two bodies are prepared in mutually independent (separable) pure quantum states with Gaussian uncertainties (also shown).
Particle A has zero momentum expectation value ($\langle \hat p_{\rm A}\!\;\!(t_0) \rangle = 0$) while its momentum uncertainty is squeezed according to $\Delta^{\!2} \hat p_{\rm A}\!\;\!(t_0) \ll \hbar m_{\rm A} \Omega/2$. Its position uncertainty is anti-squeezed according to $\Delta^{\!2} \hat x_{\rm A}\!\;\!(t_0) = \hbar^2/(4\Delta^{\!2} \hat p_{\rm A}\!\;\!(t_0))$. 
\\
Particle B's position is described by $\langle \hat x_{\rm B}(t_0) \rangle = -x_0$ with a squeezed uncertainty $\Delta^{\!2} \hat x_{\rm B}(t_0) \!\ll\! \hbar/(2 m_{\rm B} \Omega)$. It has a large positive momentum with an anti-squeezed uncertainty according to
\begin{equation}
\langle \hat p_{\rm B}(t_0) \rangle \gg \Delta \hat p_{\rm B}(t_0)\!= \!\frac{\hbar}{2 \Delta \hat x_{\rm B}(t_0)}  \gg \sqrt{\frac{\hbar m_{\rm B} \Omega}{2}}  \, .
\end{equation}
~\\
{\bf 1D-Collision with 50\% momentum transfer}\\ 
Every single collision `i' must obey momentum as well as energy conservation. 
This leads to the well-known effect that a one-dimensional elastic collision of two bodies with \emph{identical} masses swap their motional quantum states.
In this case, the two bodies remain in separable states of motion, and the collision does not produce any entanglement.\\
Keeping the 1D setting, the situation becomes different if the masses are unequal. The strongest quantum correlation occurs when the prepared bodies have a collision with 50\% momentum transfer, as the momentum uncertainties are then also transferred in the same ratio. 
The momentum transfer is both ways, but only the momentum transfer from B to A is relevant.
The other direction is irrelevant because A has zero momentum and negligible momentum uncertainty.
The transfer of momentum uncertainty is effectively `oneway'. This is what the entanglement produces.\\

The 50\% momentum transfer in Fig.\,\ref{fig:1} is described by
\begin{eqnarray} \label{eq:3a}
\frac{1}{2} m_{\rm B} v_{\rm B,i}(t_0)&\approx& m_{\rm B} v_{\rm B,i}(t_1) = p_{\rm B,i}(t_1) \\ 
							&\approx& m_{\rm A} v_{\rm A,i}(t_1) = p_{\rm A,i}(t_1) \, .
\label{eq:3b}
\end{eqnarray}
Energy conservation requires
\begin{equation}
m_{\rm B} v^2_{\rm B,i}(t_0) \approx  m_{\rm B} v^2_{\rm B,i}(t_1) +  m_{\rm A} v^2_{\rm A,i}(t_1)  \, .
\label{eq:4}
\vspace{1mm}
\end{equation}
Combining these equations provides the optimal mass ratio for maximal entanglement of
\begin{equation}
m_{\rm B} =  3\,m_{\rm A} \, .
\label{eq:5}
\end{equation}

~\\
{\bf Emergent EPR quantum correlations}\\ 
Eqs.\,(\ref{eq:3a}), (\ref{eq:3b}) and (\ref{eq:4}) are justified approximations because particle A has zero momentum and a (strongly) squeezed momentum uncertainty before the collision. No momentum and no kinetic energy is thus transferred from particle A to particle B in course of the elastic collision.
Eqs.\,(\ref{eq:3a}) and (\ref{eq:3b}) readily state that \emph{the momenta of A and B are perfectly correlated} for every individual pair collision `$i$': 
A measurement of the momentum $p_{\rm A,i}(t_1)$ allows to precisely infer the momentum $p_{\rm B,i}(t_1)$ and vice versa. 

The quantum anti-correlation in the bodies' positions can be shown by considering their velocities.
Eqs.\,(\ref{eq:3a}), (\ref{eq:3b}) and (\ref{eq:5}) yield
\begin{equation}
\frac{2}{3} v_{\rm A,i}(t_1) =  2 v_{\rm B,i}(t_1) =  v_{\rm B,i}(t_0) \equiv \langle v_{\rm B}(t_0) \rangle + \delta v_{\rm B,i}(t_0) \, , 
\label{eq:8}
\end{equation}
with $ |\delta v_{\rm B,i}(t_0) | \ll  | \langle v_{\rm B}(t_0) \rangle |$, where $\delta v_{\rm B,i}(t_0)$ can be either positive or negative describing the effect of the quantum uncertainty on individual measurement outcomes.

The time of collision $t_{\rm coll}$ has an uncertainty described by 
\begin{equation}
\delta t_{\rm coll,i} = \frac{\delta x_{\rm A,i}(t_0)}{\langle v_{\rm B}(t_0) \rangle + \delta v_{\rm B,i}(t_0)} 
\approx \frac{\delta x_{\rm A,i}(t_0)}{\langle v_{\rm B}(t_0) \rangle} \, .
\label{eq:9}
\end{equation}

As illustrated by the left circle in the figure of the supplementary information, the position of B according to a single measurement at time $t_1$ is then approximated by
\begin{eqnarray}
x_{\rm B,i}(t_1) \!&=&\! -x_0 + \left( \langle v_{\rm B}(t_0) \rangle + \delta v_{\rm B,i}(t_0) \right) \nonumber \\
&&\hspace{20mm} \times \left( \frac{3 x_0 /2}{\langle v_{\rm B}(t_0) \rangle} + \frac{\delta t_{\rm coll,i}}{2} \right) \nonumber \\
\!&\approx&\!  \frac{x_0}{2} + \frac{\langle v_{\rm B} (t_0)\rangle \delta t_{\rm coll,i} }{2} + \frac{3\delta v_{\rm B,i}(t_0)  \,x_0}{2\langle v_{\rm B}(t_0) \rangle}\nonumber \\
\!&\approx&\!   \frac{x_0}{2} + \frac{\delta x_{\rm A,i}(t_0)}{2}  \; ,  
\label{eq:10}
\end{eqnarray}
where all terms are neglected that are small compared to $\langle v_{\rm B}(t_0) \rangle \delta t_{\rm coll,i}$.\\[1mm]
The position of A at measuring time $t_1$ of run i reads
\begin{eqnarray}
x_{\rm A,i}(t_1) \!&=&\! \left( \langle v_{\rm B}(t_0) \rangle \!+\! \delta v_{\rm B,i}(t_0) \right) \nonumber  \\
&&\hspace{3mm} \times \left( \delta t_{\rm coll,i} + \frac{3 \,x_0}{2 \langle v_{\rm B}(t_0) \rangle} - \frac{3\, \delta t_{\rm coll,i}}{2} \right) \nonumber \\
\!&\approx&\!   \frac{3 \,x_0}{2} - \frac{\delta x_{\rm A,i}(t_0)}{2} \; . 
\label{eq:11}
\end{eqnarray}
Thus, it is shown that \emph{the positions of the bodies are quantum anti-correlated}. A measurement of the position $x_{\rm A,i}(t_1)$ allows to precisely infer the position $x_{\rm B,i}(t_1)$. There is no quantum uncertainty in the sum of Eqs.\,(\ref{eq:10}) and (\ref{eq:11}).

~\\
{\bf Proposal for an implementation with ions}\\
Two ions trapped in a linear Paul trap can move in one dimension and also repel each other. The experiment proposed here requires two ions of different masses, preferably in a ratio of 1:3. Potential candidates are potassium (mass number 39) and caesium (mass number 133).
In order to realise the experiment in Figure 1, the (singly charged) potassium ion (`A') must be prepared in a momentum-squeezed state before the elastic collision, and the (singly charged) caesium ion (`B') must be prepared in a position-squeezed state. This could for example be realised with superimposed, three-dimensional ion traps that have significantly different trap frequencies.
The potassium ion ($m_{\rm A}$) is initially in the ground state of a trap potential with a low trap frequency $\Omega_{\rm A}$ at the origin $\langle x_{\rm A}(t<t_0) \rangle = 0$.
The cesium ion ($m_{\rm B}$) is initially found in the ground state of a trap potential with a high trap frequency $\Omega_{\rm B} \gg \Omega_{\rm A}$ at the location $\langle x_{\rm B}(t<t_0) \rangle = -x_0$.
At $t = t_0$, both trap potentials are switched off and the linear Paul trap is activated instead, which strongly forces both ions to move in one dimension and has a trap potential of medium trap frequency $\Omega_P$ along the axis. The following therefore applies $\Omega_{\rm A} < \Omega_{\rm P} < \Omega_{\rm B}$.  Shortly after the trap frequencies $\Omega_{\rm A}$ and $\Omega_{\rm B}$ are switched off, ion A and ion B have a squeezed momentum or a squeezed position with respect to the new trapping frequency $\Omega_{\rm P}$ according to Eq.\,(\ref{eq:2}). To prevent the ion motion from being influenced by the evolution of their wave functions in the new potential, ion B is immediately accelerated towards ion A at high speed and the measurements of either the two locations or the two momenta are performed after the collision. The measurement results show the EPR pardoxon.
 
\vspace{-2mm}
\section*{Discussion}
\vspace{-2mm}
Einstein, Podolsky and Rosen concluded from their `EPR' thought experiment that quantum theory is incomplete and that the description of physical systems must be supplemented \cite{Einstein1935} by what was later called `local hidden variables'. 
However, this possibility was ruled out as part of theoretical and experimental work that began in the 1960s \cite{Bell1964,Clauser1969,Freedman1972,Clauser1978,Aspect1981,Weihs1998,Pan2000,Giustina2013,Hensen2015} and concluded with the award of the Nobel Prize in Physics in 2022.

In the original EPR thought experiment, the physical process of how the necessary entanglement comes about was missing. I can supplement this with this work. Furthermore, I visualise the generation of the entanglement of the positions and momenta of two particles through a time sequence (Fig.\,\ref{fig:1}).
A one-to-one realisation of the original EPR thought experiment thus appears possible for the first time.
As the even more important result of my work I consider the insight into how the quantum correlations of EPR entanglement arise. This follows directly from the equations and the illustration presented. EPR entanglement arises from the redistribution of the initial quantum uncertainties under the conditions of energy and momentum conservation. If the initial quantum uncertainties are position- or momentum-squeezed, the redistributions are effectively one-way streets, and quantum correlations and entanglement arise in an easily understandable way. 

The generation and measurement of EPR entanglement in this paper refers exclusively to the statement of the original EPR thought experiment, which can be adequately described by quantum states with Gaussian (positive) Wigner functions.  
In general, Gaussian entanglement together with the measurement of Gaussian variables (here position and momentum) is not suitable for violating a Bell inequality, since this situation can be described with a local deterministic model \cite{Bell1987}. (This is only possible if one disregards the existence of Heisenberg's uncertainty principle).
However, Gaussian entanglement can make a locally deterministic model entirely impossible if non-Gaussian variables are measured, in particular parity \cite{Banaszek1998,Royer1977}.
In the context of quantum information, it is a very interesting question whether it is possible to define parity measurements on the Gaussian EPR entangled states of this paper in such a way that the measurement results violate a Bell inequality. This question is not answered here, but I suspect that it is possible.

\vspace{-2mm}
\subsection*{Data Availability}
\vspace{-4mm}
Not applicable.

\vspace{-2mm}
\subsection*{Acknowlegement}
\vspace{-4mm}
This work was partially performed within the European Research Council (ERC) Project \emph{MassQ} (Grant No.~339897) and within Germany's Excellence Strategy -- EXC 2056 `Advanced Imaging of Matter', project ID 390715994 and EXC 2121 `Quantum Universe', project ID 390833306, which are financed by the Deutsche Forschungsgemeinschaft. The author thanks Jesse Everett for pointing out that the original illustrations did not match the equations and were wrong.

\vspace{-2mm}
\subsection*{Competing Interests}
\vspace{-4mm}
The author declares no competing interests.

\newpage
\vspace{3cm}
\newpage

{\Large{\bf --- Supplementary Information ---}}

\section{Illustrating the entangling collision in phase space}
The physical content of the figure of the main manuscript can alternatively be represented in the phase space of positions and momenta, see Fig.\,\ref{fig:1s} below. Such a phase space representation corresponds to the usual quantum mechanically complete representation.\\

\begin{figure*}[h!!!!!!!!!!!!!!!!!!!!!!!!!!!!!!!!!!!!!!!!!!!!!!!!!!!!!!!!!!!!!!!!!!!!!!!!!!!!!!!!!!!!!!!!!!!!!!!!!!!!!!!!!!!!!!!!!!]
    \vspace{0mm}
\includegraphics[width=14cm]{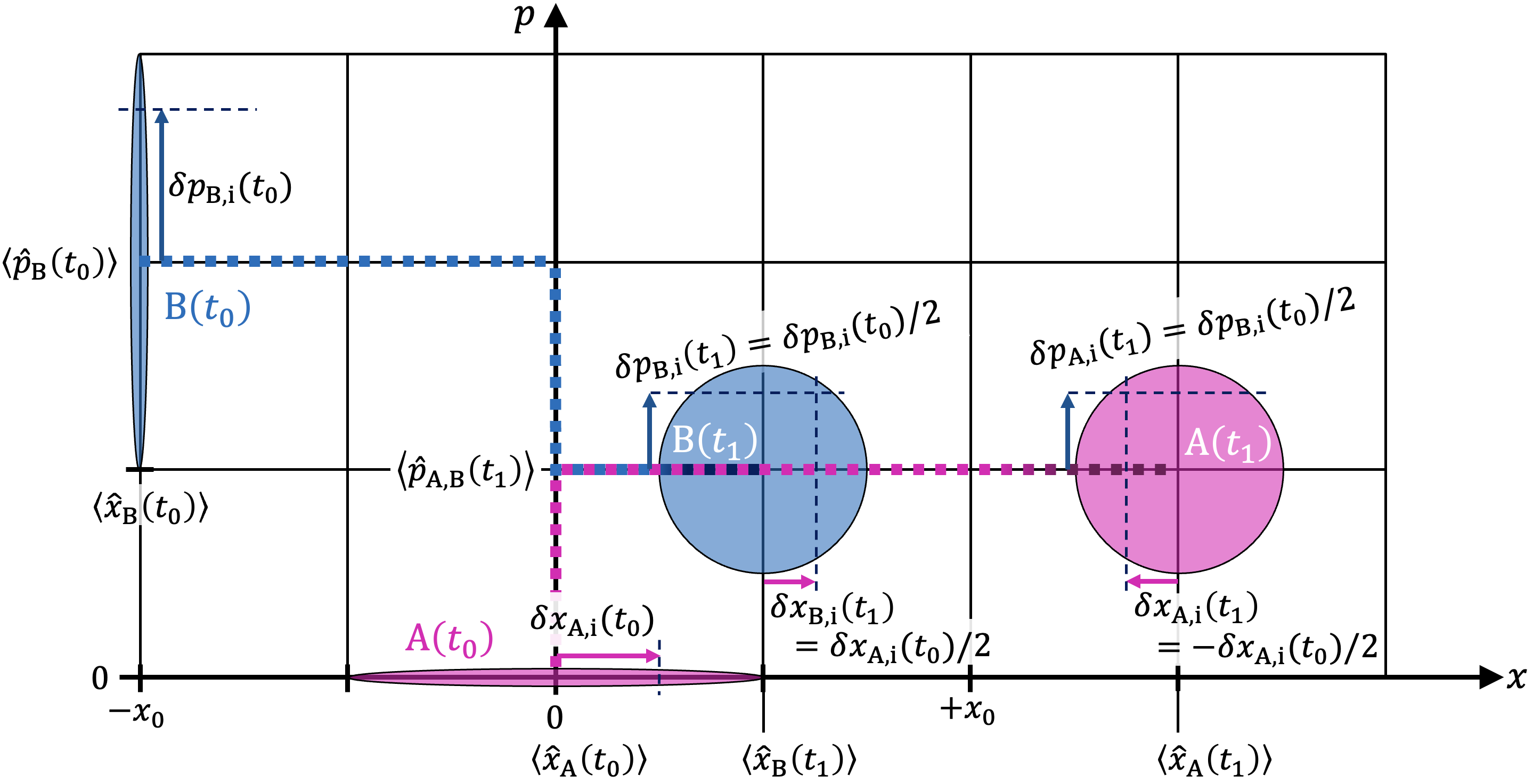}
    \vspace{-1mm}
    \caption{{\bf EPR entanglement from conventional elastic collision in phase space} -- Shown are expectation values surrounded by Gaussian quantum uncertainties (thin uncertainty areas) of the positions and momenta of two bodies A and B at time $t_0$ shortly before their collision at position $\langle x_{\rm A}(t_0)\rangle$, as well as at time $t_1$ shortly after their collision (large uncertainty areas). The two thin ellipses represent squeezed, mutually uncorrelated minimal uncertainties before the collision. The two circular uncertainty areas represent EPR entangled uncertainties after the collision. Individually, they are not minimal. The entanglement-generating collision happens at  time $\langle \hat t_{\rm coll} \rangle$ with $t_0 \!<\! \langle \hat t_{\rm coll} \rangle \!<\! t_1$. (The calculation in the main text shows that the time of the collision must actually be described by an ``arrival time'' operator [1,2].) The bodies' mass ratio is $m_{\rm B} = 3 m_{\rm A}$, due to which the initial momentum of B is equally distributed amongst A and B. The distribution of the momentum of B is independent of its magnitude, from which it follows that the momentum \emph{uncertainty} of B is also distributed equally. With the approximation that the initial momentum uncertainty of A is negligible, the momentum uncertainties after the collision are perfectly quantum correlated. Subscript `i' represents a single measurement example. 
    The positions after the collision are determined as follows: 
    B's velocity is halved, while A's velocity is three-times larger than that of B. This determines the position expectation values at time $t_1$ as shown. The large initial position uncertainty of A, however, creates a collision timing uncertainty. If the collision already takes place when B reaches the left end of A's ellipse of uncertainty, body A gains speed earlier and is at the right end of the circular uncertainty at time $t_1$. In contrast, the body B has its reduced velocity earlier and ends up at the left end of its uncertainty circle. The spatial uncertainty after the collision are perfectly quantum-anticorrelated, if one again makes the approximation that the initial squeezed spatial uncertainty of B is negligible. 
    \emph{The figure makes obvious why EPR correlations are observed. 
    By measuring either A or B we can predict with certainty, and without in any way disturbing the second system, either the value of $\,x\,$ or the value of $\,p\,$.}\\[3mm] 
$[1]$ J. Kijowski, ``On the time operator in quantum mechanics and the Heisenberg uncertainty relation for energy and time'', {\em Reports on Mathematical Physics}, vol.~6, 361--386, 1974.\\[3mm] 
$[2]$ J. Kiukas, A. Ruschhaupt, P. O. Schmidt, and R. F. Werner, ``Exact energy--time uncertainty relation for arrival time by absorption'', {\em J. Phys. A: Math. Theor.}, vol.~45, 185301, 2012.
    }
    \label{fig:1s}
    \vspace{2mm}
\end{figure*}

\section{Mass ratio 1:3}
Case (i): The lighter mass (`A') initially has zero momentum and a strongly squeezed momentum uncertainty.
Here, every collision `i' of an ensemble of identical experiments with 50\% momentum transfer is described by
\begin{equation} \nonumber
\frac{1}{2} m_{\rm B} v_{\rm B,i}(t_0) =  m_{\rm B} v_{\rm B,i}(t_1) = m_{\rm A} v_{\rm A,i}(t_1)  \, .
\end{equation}
Energy conservation enforces
\begin{equation} \nonumber
m_{\rm B} v^2_{\rm B,i}(t_0) =  m_{\rm B} v^2_{\rm B,i}(t_1) +  m_{\rm A} v^2_{\rm A,i}(t_1)  \, ,
\end{equation}
from which follows
 \vspace{-3mm}\begin{equation} \nonumber
m_{\rm B} =  3\, m_{\rm A} \, .
\end{equation}

Case (ii): The heavier mass `A' initially has zero momentum and a strongly squeezed momentum uncertainty.
Here, every collision `i' of an ensemble of identical experiments with 50\% momentum transfer involves a momentum sign flip and is described by
\begin{eqnarray} \nonumber
\frac{1}{2} p_{\rm B,i}(t_0) &\!=\!&  - p_{\rm B,i}(t_1) \, , \nonumber \\
p_{\rm B,i}(t_0) &\!=\!& p_{\rm B,i}(t_1) + p_{\rm A,i}(t_1) \, , \nonumber \\
\Rightarrow \;\;\; \nonumber
\frac{1}{2} m_{\rm B} v_{\rm B,i}(t_0) &\!=\!&  - m_{\rm B} v_{\rm B,i}(t_1) = \frac{1}{3} m_{\rm A} v_{\rm A,i}(t_1)  \, .\;\;\;\;\;
\end{eqnarray}
Energy conservation enforces in this case
\begin{equation} \nonumber
m_{\rm B} =  \frac{1}{3} m_{\rm A} \, .
\end{equation}
A one dimensional elastic collision between two particles with mass ratio 1:3 results in a 50\% momentum transfer.
It corresponds to a balanced redistribution of momentum expectation values as well as momentum uncertainties. 
The redistribution happens in both directions, from `A' to `B' and vice versa.
However, if one particle has negligible (strongly squeezed) momentum uncertainty, the momentum transfer is de facto one way.
Such a collision produces maximal quantum correlation of the particles' momenta if the mass ratio is 1:3.

\end{document}